\documentclass[oldversion, preprint]{aa}  
\usepackage{graphicx}
\usepackage{txfonts}
\usepackage{longtable}
\begin{document}
   \title{Radio emission of SN1993J: The complete picture.}
   \subtitle{II. Simultaneous fit of expansion and radio light curves.}

   \author{I. Mart\'i-Vidal\inst{1,2}
          \and
          J.M. Marcaide\inst{2}
          \and 
          A. Alberdi\inst{3}
          \and
          J.C. Guirado\inst{2}
          \and 
          M.A. P\'erez-Torres\inst{3}
          \and
          E. Ros\inst{2,1}
          }

   \institute{Max-Planck-Institut f\"ur Radioastronomie,
             Auf dem H\"ugel 69, D-53121 Bonn (Germany)
          \and 
              Dpt. Astronomia i Astrof\'isica, Universitat de Val\`encia,
              C/ Dr. Moliner 50, E-46100 Burjassot (Spain)\\
              \email{imartiv@mpifr-bonn.mpg.de} 
         \and
             Instituto de Astrof\'isica de Andaluc\'ia (CSIC),
             C/ Camino bajo de Hu\'etor 50, E-18008 Granada (Spain)
             }

   \date{Accepted for publication in A\&A.}
 
  \abstract
{
We report on a simultaneous modelling of the expansion and radio light 
curves of the supernova SN1993J. We developed a simulation code capable of generating 
synthetic expansion and radio light curves of supernovae by taking into 
consideration the evolution of the expanding shock, magnetic fields, and 
relativistic electrons, as well as the finite sensitivity of the interferometric 
arrays used in the observations. Our software successfully fits all the available 
radio data of SN 1993J with a standard emission model for supernovae, which is extended with 
some physical considerations, such as an evolution in the opacity of the ejecta material, a 
radial decline in the magnetic fields within the radiating region, and a changing radial 
density profile for the circumstellar medium starting from day 3100 after the explosion. 
}

   \maketitle

\section{Introduction}
\label{I}

We previously reported on an analysis of the complete set of available VLBI
observations of SN\,1993J (Mart\'i-Vidal et al. \cite{MartiVidal2009}, hereafter 
Paper I). In that work, we confirmed the main results reported in Marcaide et al.
(\cite{Marcaide2009}) about an expansion curve that is dependent on the observing frequency.
These results are not compatible with those published by other authors 
(Bartel et al. \cite{Bartel2002}), who claimed up to four different values 
of the expansion index, $m$ (where the supernova radius is $R \propto t^m$ and $t$ is 
the age after explosion, see Chevalier \cite{Chevalier1982a}), corresponding to four 
different expansion periods. The interpretation of the 
data reported in Marcaide et al. (\cite{Marcaide2009}) was very different: 
there is essentially one expansion index during the whole supernova expansion 
(in Paper I, we report however on two expansion regimes separated by an early 
{\em break time} at $t\sim390$ days). Two explanations were then proposed in 
Marcaide et al. (\cite{Marcaide2009}) for the different shell sizes found at 
different frequencies. 
On the one hand, an evolution in the opacity of the ejecta to the radio emission. 
This opacity was assumed to be maximum (i.e., 100\%) at the lowest frequency (1.7\,GHz) 
and slowly decrease in time at the higher frequencies, beginning about day 1500 
after the explosion. On the other hand, a radial decay in the amplified 
magnetic fields within the emitting shell was also proposed.
This profile in the magnetic-field distribution translates 
into a profile in the emission intensity, which (combined with the finite sensitivity 
of the interferometers used in the observations) can also help us to explain the expansion 
curve. In this paper, we quantify the effects proposed in Marcaide et al. 
(\cite{Marcaide2009}) and Paper I by developing a new software capable to modelling 
simultaneously the expansion and radio light curves of SN\,1993J.

The radio light curves of SN\,1993J were previously modelled using
several approaches: Weiler et al. (\cite{Weiler2002}, \cite{Weiler2007}) used an 
analytical model to fit the data; Fransson \& Bj\"ornsson (\cite{Fransson1998}) 
and P\'erez-Torres et al. (\cite{PerezTorres2001}) simulated the
evolution of the relativistic electron population inside the radiating region 
by taking into account the hydrodynamics of the shock evolution described in 
Chevalier (\cite{Chevalier1982a}) and the radiative cooling of the
relativistic electrons; finally, Chandra, Ray \& Bhatnagar (\cite{Chandra2004}) took into 
account synchrotron-ageing effects 
on the electron population. The claim of a steeper spectral index at 
high frequencies and late epochs made by Chandra, Ray \& Bhatnagar (\cite{Chandra2004}) 
contradicted previous reports (P\'erez-Torres, Alberdi, \& Marcaide 
\cite{PerezTorres2002}; Bartel et al. \cite{Bartel2002}) and was not confirmed by 
Weiler et al. (\cite{Weiler2007}), who reported instead a 
flattening of the spectral index of the supernova at late epochs and all frequencies.

The remainder of this paper is structured as follows. In Sect. \ref{II}, 
we describe our new code for the simultaneous modelling 
of the SN\,1993J radio light curves and expansion curve. In Sect. \ref{III}, 
we report on the final fitted model and describe the modifications (or 
extensions) of the Chevalier model (Chevalier \cite{Chevalier1982a}, 
\cite{Chevalier1982b}) that were applied to our software to 
properly model the whole data set. In Sect. \ref{V}, we summarize our 
conclusions.

\section{RAMSES: a simulator of the synchrotron emission from 
supernovae}
\label{II}

To fit the radio light-curve data and, at the same time, fit the expansion curve of 
SN\,1993J, we developed a new simulation code, RAMSES ({\em Radiation-Absorption 
Modeller of the Synchrotron Emission from Supernovae}). A detailed description 
of much of this model can be found in Mart\'i-Vidal (\cite{MartiVidal2008}). 
Some relevant aspects of the algorithms implemented in the program are also described in Appendix 
\ref{RAMSESApp} of this paper. In this section, we summarize the main characteristics 
of RAMSES, which is based on the work of Chevalier (\cite{Chevalier1982a},\cite{Chevalier1982b}) 
and extended in several aspects. It assumes that: (1) the expansion is self-similar; (2) 
the radio-emitting region of the supernova is located between the contact discontinuity and 
the forward shock (the width of this spherical shell being 30\% of the shell radius, see 
Marcaide et al. \cite{Marcaide2009} and Paper I); (3) a given fraction, $f_{rel}$, of the 
thermal electrons of the CSM are accelerated by the shock to relativistic energies and 
characterized by an energy distribution $\propto E^{-p}$, where $p$ is the {\em energy index} 
of the electron population; (4) the electrons emit synchrotron radiation as they interact with 
an amplified magnetic field that fills the shocked circumstellar region; (5) the mean intensity 
of the magnetic field depends on the distance from the contact discontinuity; (6)
the radius of the spherical surface defined by the contact discontinuity (although without 
considering the development of Rayleigh-Taylor fingers at this point) expands as $R \propto t^m$, 
where $m$ is the {\em deceleration parameter} or expansion index ($m = (n - 3)/(n - s)$); (7) 
during the expansion, the opacity of the ejecta to the radio emission may also evolve, and be 
different for different frequencies. 

Our code RAMSES considers radiative cooling, inverse Compton scattering, and synchrotron 
self-absorption (SSA) in both the electron energy distribution and the radio emission. It 
generates synthetic images of the supernova that provide
predictions of the expansion and radio light curves of SN\,1993J. To determine 
the expansion curve, a high-pass filter of the flux density per unit beam is initially applied to 
the synthetic images, depending on the sensitivity of the VLBI arrays used in the observations. 
The Common-Point Method is then applied to the resulting images as in the case of real data (see 
Marcaide et al. \cite{Marcaide2009}) to obtain the model of the VLBI expansion curve. 

Several parameters are fitted to the observed data by RAMSES (see Table \ref{RAMSESTable} for a 
summary), including the: (1) density of CSM electrons, $n_{csm}$, in the emitting region at a given 
(reference) epoch; (2) mean magnetic field intensity, $\bar B$ (i.e., average of the magnetic 
field intensity 
over the emitting region) at the same epoch; (3) energy index, $p$, of the electron energy 
distribution; (4) fraction, $f_{rel}$, of the CSM electrons that are accelerated by the shock 
(i.e., the acceleration efficiency of the shock); (5) radial profile of the temperature of the 
unshocked CSM (which affects the free-free absorption, FFA, of the CSM, especially at early 
epochs, see Appendix \ref{RAMSESApp}); and (6) mean lifetime, $t_{m}$, of the relativistic 
electrons inside the supernova shell before they escape outside the emitting region. In 
Appendix \ref{RAMSESApp}, we explain in more detail the 
meaning of these parameters. RAMSES assumes a power-law time evolution for the magnetic-field
energy density and the energy density of the electrons at the shock, and generates the synthetic 
light curves and expansion curve. The model is then fitted to the observations by means of a 
least squares minimization (see Appendix \ref{RAMSESApp} for more details). The expansion curve 
is parametrized in the same way as in Paper I (i.e., using two expansion indices, $m_1$ and 
$m_2$, separated by a break time, $t_{br}$). Thus, our fits use 9 parameters. The ``Best Fit'' 
(SSA + FFA) model reported in Weiler et al. (\cite{Weiler2007}) to fit the radio-light curves 
also had 9 fitting parameters, even though these authors did not include the expansion curve in 
their fits. In addition, we adopt several {\em ad hoc} assumptions about the evolution of the 
ejecta opacity, the profile of the CSM density, and the profile of the magnetic field, as 
described in the next few sections.

\begin{table*}
\centering
\begin{tabular}{ c | c | c }
\hline\hline
\multicolumn{3}{c}{   } \\
\multicolumn{3}{c}{\bf {Fitted parameters}} \\
\hline
$m_1 = 0.925 \pm 0.016$ & $\bar{B}_0 = 65.1 \pm 1.6$ G           & $n_0 = (6 \pm 0.9)\times 10^8$ cm$^{-3}$ \\
$m_2 = 0.87 \pm 0.02$   & $f_{rel} = (5 \pm 0.5)\times 10^{-5}$  & $T_l = (2.0\pm0.1)\times 10^6$ K \\
$t_{br} = 360 \pm 50$   & $p=-2.55 \pm 0.01$                     & $t_m = 2500 \pm 100$ days  \\
\hline
\multicolumn{3}{c}{   } \\
\multicolumn{3}{c}{\bf{Assumptions and fixed parameters}} \\
\hline
$ s = 2 $        & $\bar{B}^2 \propto n_{cs}\,V^2$   & Changing ejecta opacity (vid. Fig. \ref{Opac}, left) \\
$t_0 = 5.3$ days & $n_{rel}\,dt \propto n_{cs}\,V^2$ & Non-uniform $B$ (vid. Fig. \ref{Opac}, right) \\
$ D_{M\,81} = 3.63$ Mpc & $t_{expl} = $ 1993 March 28 &  \\
\hline
\end{tabular}
\caption{Summary of fitted parameters, fixed parameters, and assumptions in the 
RAMSES model. Details on the meaning of these quantities are given in Sect. 
\ref{II} and Appendix \ref{RAMSESApp}.}
\label{RAMSESTable}
\end{table*}

\section{Simultaneous fit of expansion and radio-light curves}
\label{III}

\begin{figure*}[!t]
\centering
\includegraphics[width=15cm]{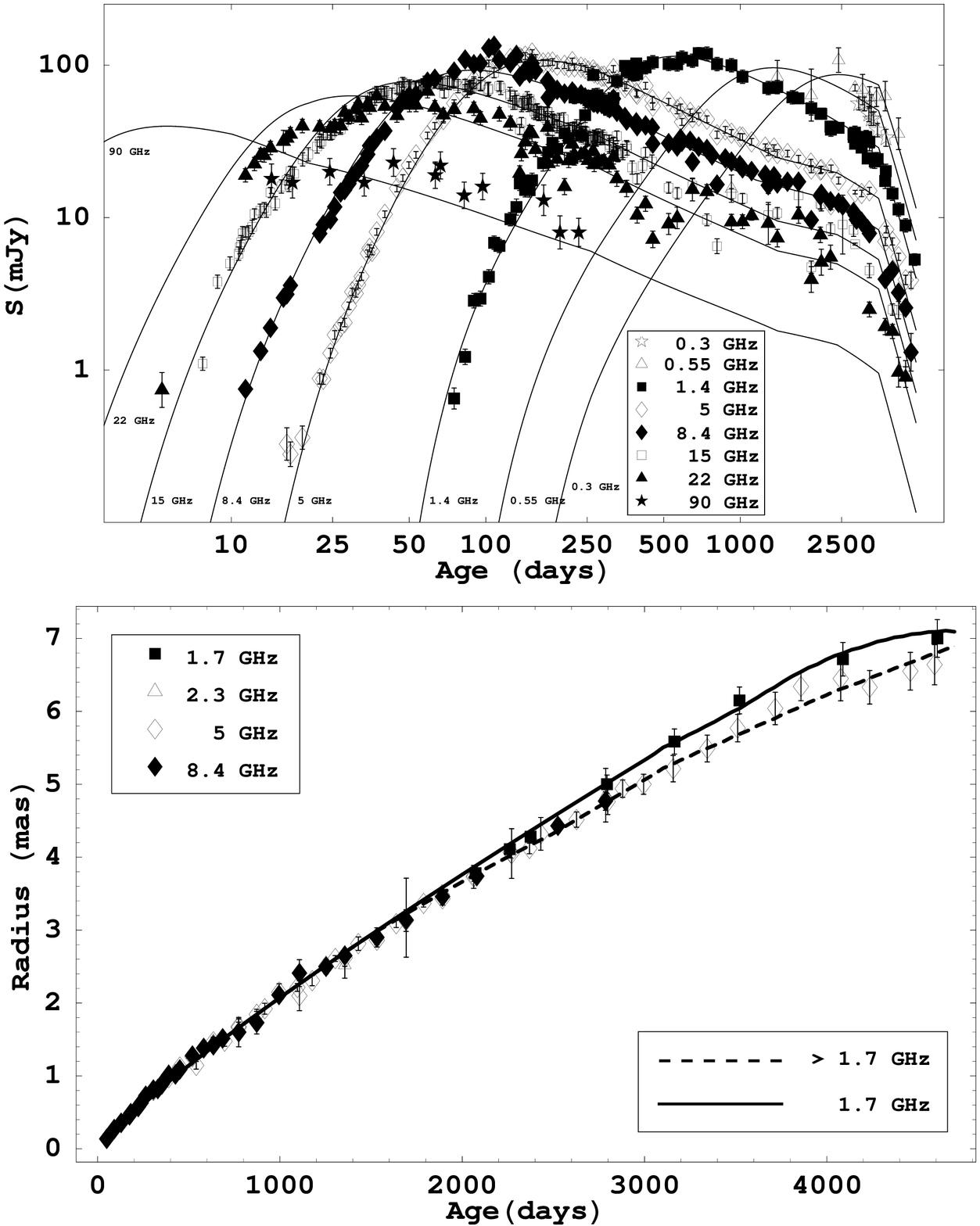}
\caption{Top, fit of RAMSES to the radio light curves reported in
 Weiler et al. (\cite{Weiler2007}) up to an age of 4930 days. Bottom, 
simultaneous fit of RAMSES to the expansion curve reported in Paper I.}
\label{SIMULFIT}
\end{figure*}

In Fig. \ref{SIMULFIT}, we show the fit of the RAMSES model to the radio 
light curves reported in Weiler et al. (\cite{Weiler2007}) (top) and to the 
expansion curve reported in Paper I (bottom). 
In the fit, the structure index of the CSM was fixed to $s = 2$ 
and the CSM density was set to be 0 after day 3100 (see Sect. \ref{NOCSM}). 
For the reference epoch taken at 5.3 days after the 
explosion (when the radius of the contact discontinuity of the spherical shell is 
$10^{15}$\,cm), we obtain a minimum reduced-$\chi^2$ of 5.4 for the following values of the 
fitting parameters (see Table \ref{RAMSESTable} for a summary): mean magnetic field, 
$\bar B$, $65.1\pm1.6$\,G; post-shock circumstellar 
electron number density, $n_{csm}$, $(6 \pm 0.9)\times 10^{8}$ cm$^{-3}$; 
acceleration efficiency, $f_{rel}$, $(5 \pm 0.5)\times 10^{-5}$;
energy index of the electron energy distribution, $p$, $-2.55\pm0.01$; and a mean lifetime 
of the electrons inside the shell, $t_{m}$, $2500 \pm 100$\,days.
The fitted $\bar B$, $n_{csm}$, $f_{rel}$, and $p$ are similar to 
those reported by Fransson \& Bj\"ornsson (\cite{Fransson1998}).
For values of $s$ lower than 2 (i.e., $s = 1.6 - 1.7$), it is impossible 
to obtain satisfactory simultaneous fits to radio light curves and the expansion curve.

The best-fit model parameters for the expansion curve are: $m_1 = 0.925 \pm 0.016$, 
$m_2 = 0.87 \pm 0.02$, and  $t_{br} = 360 \pm 50$\,days. The fitted $m_2$ is practically the same 
expansion index as that reported in Paper I for the shell sizes at 1.7\,GHz ($m_3$ in that 
paper), and is close to the index reported in Marcaide et al. (\cite{Marcaide2009}) ($m_1$ in that 
paper). The lower expansion index previously reported for the higher-frequency data (which 
corresponds to  $m_2$ in Paper I and in Marcaide et al. \cite{Marcaide2009}) is modelled by 
RAMSES by considering several physical and instrumental effects (outlined in Marcaide et al. 
\cite{Marcaide2009}) that we describe in the next few subsections.

\subsection{Particle-field energy equipartition}

Our model assumes that the energy transmitted to the accelerated electrons by the shock at 
each time is proportional to the energy of the magnetic field, which is in turn proportional
to the energy of the shock (see Table \ref{RAMSESTable}). However, the proportionality 
constants between these quantities are not fixed in our model, since we fit simultaneously 
the particle number density, the magnetic field (both at a reference epoch), and the 
acceleration efficiency without any a priori covariance. Therefore, particle-field energy 
equipartition is not assumed in our model. However, we can estimate the level of energy 
equipartition between particles and fields from our best-fit model
(an approach similar to that of Fransson \& Bj\"ornsson \cite{Fransson1998}):

On the one hand, the energy density of relativistic electrons, relative to that of the 
thermal electrons in the shock, runs smoothly from $\sim 6\times10^{-5}$ at early epochs to
$3-4\times10^{-4}$, at late epochs. If the acceleration efficiency of the ions were similar
to that of the electrons, we would reach a rough level of equipartition between thermal and 
non-thermal particles in the shock.

On the other hand, the energy density of the magnetic field, relative to that of the thermal
electrons, runs smoothly from $0.25$ at early epochs to $0.06$ at late epochs. In this case,
the level of equipartition is not large, the magnetic field being a factor $2-4$ below the 
equipartition value with the thermal shocked electrons.

Another way to analyze the level of equipartition is to compute the energy
density of the magnetic field relative to that of the accelerated particles. In this case, the 
ratio runs from $\sim$3000 at early epochs to $\sim$170 at late epochs. Therefore, 
equipartition between magnetic field and relativistic particles is roughly obtained,
again, provided that the efficiency in the acceleration of ions is similar to that of the 
electrons.


\subsection{Model of the CSM}

\subsubsection{FFA versus SSA} 

\begin{figure}
\centering
\includegraphics[width=9cm]{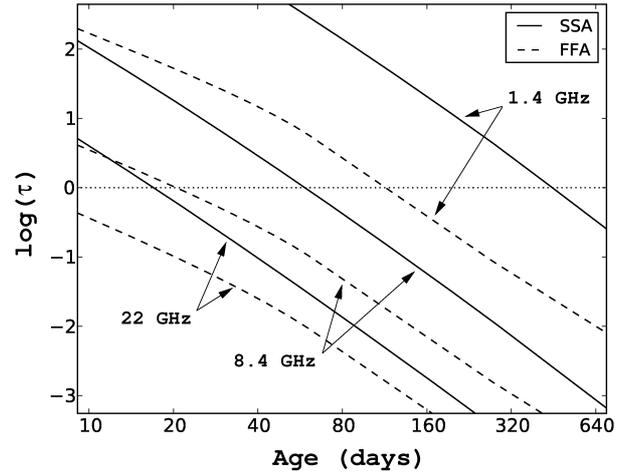}
\caption{SSA opacities (solid lines) and FFA opacities (dashed lines) versus time for a selection
of frequencies.}
\label{FFAvsSSA}
\end{figure}

In Fig. \ref{FFAvsSSA}, we show the SSA and FFA opacities computed with RAMSES in the 
optically-thick part of the light curves at 1.4, 8.4, and 22\,GHz. It can be seen in the 
figure that SSA dominates versus FFA for all frequencies and epochs. The FFA evolution 
changes its slope at $\sim$80\,days, because of the radial gradient of unshocked electron 
temperatures used in the model of the CSM (see next section and Appendix A). 
It can also be seen in Fig. 
\ref{FFAvsSSA} that, for each frequency, the opacity of SSA is $\tau \sim 1$ at 
roughly the same time as the light curve reaches its maximum (see also Fig. \ref{SIMULFIT}), 
as expected for SSA-dominated light curves. Chevalier (\cite{Chevalier1998}) 
studied the case of SSA-dominated light curves in radio supernovae and reported a relationship 
between the maximum flux-density at a given frequency, the supernova age at the maximum, and 
the mean expansion velocity of the shock. The mean expansion velocity 
estimated in this way is a lower bound to the true shock velocity when FFA is not negligible 
relative to SSA. The ratios of VLBI-inferred expansion velocities to those estimated using the 
relationship of Chevalier (\cite{Chevalier1998}) can be computed for several radio 
supernovae: this ratio is $\sim$2.6 for SN\,1986J and SN\,1979C (from the results 
reported in P\'erez-Torres et al. \cite{PerezTorres2002b}, Bartel \& Bietenholz 
\cite{Bartel2003}, and Marcaide et al. \cite{Marcaide2009b}), $2.3-3.6$ for 
SN\,2004et (Mart\'i-Vidal et al. \cite{MartiVidal2007}), and $\sim$2.5 for SN\,2008iz 
(Brunthaler et al. \cite{Brunthaler2010}). However, for SN\,1993J we find a much smaller 
ratio ($\sim$1.5), which is indicative of a low FFA (or a high level of SSA 
compared to FFA), thus supporting the model here proposed for the SN\,1993J radio emission.

We note that the minimum energy in the population 
of accelerated electrons (which we set to $m_e\,c^2$) affects the SSA opacity and, 
therefore, plays a role in the estimate of the electron number density,
$n_{csm}$. If a low-energy cutoff is applied to the electron population, SSA decreases, 
although the effects of this 
cutoff are only significant at very early epochs, when the magnetic field is high and the 
electron energies corresponding to high critical frequencies are low. 
In the optically-thin part of the light curves, a low-energy cutoff does not change the 
emissivity, but when the source is optically thick a low-energy cutoff implies a larger 
source function (i.e., $\epsilon_\nu/\kappa_\nu$) and therefore a larger fitted $n_{csm}$ 
(to raise FFA in the optically-thick region and maintain the fit of the model to the 
observations). This, in turn, slightly decreases the fitted magnetic field, to fit 
the model to the observations in the optically-thin part of the light curves. 
In short, the use of a low-energy cutoff increases the importance of FFA relative to SSA 
(for instance, by a factor $\sim$2 if a cutoff of $10\,m_e\,c^2$ is applied), but the overall
effects in the model light curves is very small.

\subsubsection{Evolution of FFA opacity in the CSM}

\begin{figure}[!t]
\centering
\includegraphics[width=9.5cm]{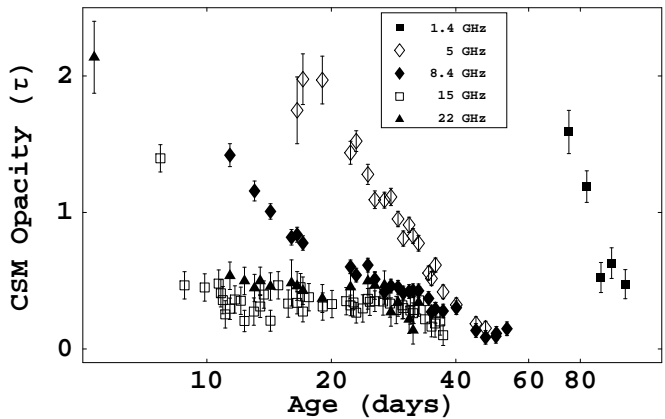}
\caption{Integrated line-of-sight free-free opacity of the unshocked CSM at several
frequencies for the early phase of the supernova expansion, measured to be the ratio 
of observed flux densities to a SSA model.}
\label{CSMTemp}
\end{figure}

The flux-density evolution of a supernova in the optically-thick regime can provide detailed
information on the radial density profile of the CSM. The opacity of the unshocked CSM to the
radio emission is assumed to be produced by free-free interaction of the radiation with the 
thermal electrons that fill the ionized CSM. Therefore, a precise estimate of the CSM temperature 
is required to reliably model the CSM density profile using the radio light curves. 
By analysing the early X-ray and radio light curves of SN\,1993J at several frequencies, 
Fransson, Lundqvist \& Chevalier (\cite{Fransson1996}) modelled a CSM radial density profile with an 
structure index $s \sim 1.7$, which is indicative of variable mass-loss rate for the precursor 
star. Similar
results were found from the analysis of the late X-ray emission (Immler, Aschenbach \& Wang 
\cite{Immler2001}). 
However, Fransson \& Bj\"ornsson (\cite{Fransson1998}) were able to model the early radio light curves 
with $s = 2$ by adding cooling effects to the electron population and assuming a dependence of the 
CSM temperature on distance from the explosion centre, which maps into a radial dependence of the 
FFA. The use of this extra degree of freedom in the model resulted in a satisfactory fit to the 
radio light curves without assuming a variable mass-loss rate of the precursor star. In our work, 
we adopted the same approach as Fransson \& Bj\"ornsson (\cite{Fransson1998}). However, the
added constraints provided by the expansion curve in the simultaneous
fit made it difficult to obtain a good fit to the 
high-frequency (22\,GHz) light curves at the earliest epochs (see Fig. \ref{SIMULFIT}). Our model 
overestimates 
the flux densities at 22\,GHz in the optically-thick regime. However, the true radial 
profile of the CSM temperature could behave in many ways close to the explosion 
centre and/or evolve with time after the ionization produced by the initial flare from the 
supernova. This might explain the systematics in the 22\,GHz residuals of the radio-light curve. 
A superior approach to using a model for the CSM opacity is possibly to estimate the opacity 
observationally, by comparing the measured flux densities to the opacity-free RAMSES 
predictions. In Fig. \ref{CSMTemp}, we show our estimate of the evolution of the CSM free-free 
opacity at different frequencies. The flux density of the supernova was reproduced by the 
RAMSES model without FFA, multiplied by $\exp{(-\tau)}$, where $\tau$ is the opacity 
shown in Fig. \ref{CSMTemp}. 

We note that it is impossible to model the data shown in Fig.
\ref{CSMTemp} using a simple model for the CSM temperature, as that described in Appendix 
\ref{RAMSESApp}.
The presence of inhomogeneities (i.e., clumps) in the CSM of a given radial distribution could 
help us to model the data. Indeed, the two earliest data points (the first one at 22\,GHz and the 
second one at 15\,GHz) in Fig. \ref{CSMTemp} do not follow the same general trend as the remainder 
of the data. 
The opacity at these two epochs is larger than expected from the backward extrapolation of the 
general 
trends. A possible explanation of these large opacities at very early epochs (earlier than 10 days 
after 
explosion) could be the presence of strong inhomogeneities (clumps) in the CSM close to the explosion
centre. A rapid evolution of the CSM temperature to explain these large opacity changes is less realistic, 
since it would imply a sudden {\em re-heating} of the CSM during the first days after the shock 
breakout.

\subsubsection{CSM radial density profile}

The break time fitted around day 360 after explosion may be due to an evolution in the structure
index, $n$, of the ejected material, to a change in the structure index, $s$, of the CSM,
or to a combination of both. In any case, if a value $s = 2$ is assumed after this early break, 
the parameter $n$ after the break takes an effective value of 9.7, according to the 
Chevalier model. This relatively low value of $n$ implies that there has been an enhancement of 
X-ray luminosity originating in the shocked ejecta region (although the X-ray emission could 
still be dominated by 
the circumstellar shock) given that the shock is more adiabatic and, therefore, its opacity 
becomes smaller (see Fransson, Lundqvist \& Chevalier \cite{Fransson1996}). When Fransson, 
Lundqvist \& Chevalier (\cite{Fransson1996}) and Immler, Aschenbach \& Wang (\cite{Immler2001}) 
estimated that $s \sim 1.6 - 1.7$ from their X-ray data, they did not consider the effect of a 
greater X-ray luminosity from the shocked ejecta due to $n \sim 10$.
Mioduszewski, Dwarkadas \& Ball (\cite{Miodus2001}) simulated radio images and the radio 
light curves of SN\,1993J and also claimed that $s \sim 1.7$ provides the best 
fit to the data, although these authors did not take the electron radiative cooling into 
account. More recently, Nymark, Chandra \& Fransson (\cite{Nymark2009}) reported on a 
fit to the SN\,1993J X-ray data using a model with $s=2$ and an X-ray emission dominated by 
the reverse shock. Chandra et al. (\cite{Chandra2009}) were also successful in modelling 
the X-ray data using $s=2$.

It is worth noticing that the wide fractional shell reported in Marcaide et al. 
(\cite{Marcaide2009}) and Paper I ($\sim$30\% of the outer radius) is incompatible with 
$s < 2$ in the frame of the Chevalier model (Chevalier \cite{Chevalier1982a}), because 
for $s = 2$ this shell implies that $n \sim 6$ (see Table 1 of Chevalier \cite{Chevalier1982a}). A 
lower value of $s$ would imply an even lower value of $n$, which must be larger 
than 5 for a self-similar expansion (Chevalier \cite{Chevalier1982a}). On the other hand, the 
combination of $n \sim 6$ and $s = 2$ translates into $m = 0.75$, a value much 
smaller than $m = 0.87$ (our fitted value after the break) although closer to 
the expansion index fitted to the 5\,GHz data ($m \sim 0.8$, see Marcaide et al. \cite{Marcaide2009} 
and Paper I), which, in our interpretation, does not describe the true supernova expansion. 
Could the similarity between the expansion index at 5\,GHz and the theoretical value derived from 
a 30\% fractional shell-width indicate that the true expansion curve (i.e., that corresponding 
to the forward shock) is traced by the high-frequency data? In this case, the evolution of the ejecta 
opacity would have been the opposite of that proposed in Marcaide et al. (\cite{Marcaide2009}) and Paper I. 
That is, the ejecta would have been transparent to the radio emission at all frequencies and early epochs, 
and would have become increasingly opaque to the 1.7\,GHz radiation after day 1500. We rule out this 
possibility, since in this case the fit to the radio light curves would be poorer and a value of $n$ close to 6 
(indeed, close to the limiting value $n = 5$ to keep the expansion self-similar) be too low. Baron et al. 
(\cite{Baron1995}), for instance, fitted ejecta density profiles with $n \sim 10$ for SN\,1993J, based on 
non-local thermodynamic equilibrium (NLTE) synthetic spectra. The wide fractional shell reported in 
Marcaide et al. (\cite{Marcaide2009}) and Paper I therefore remains unexplained\footnote{We note that even the narrower shell (25\% of the outer radius) reported in Bietenholz et al. 
(\cite{Bietenholz2003}) translates into a low value of $n$. In this case, we have $n \sim 7$ for 
$s = 2$. This estimate of $n$ would be even lower if $s < 2$.}.
 
In any case, we notice that it is difficult to discern whether $s$ takes the value of 2 or 
smaller, since there is a 
tight coupling between the different variables that affect the evolution of the radiation produced by 
the circumstellar interaction of the expanding supernova shock. Our software satisfactorily 
fits the radio data assuming $s = 2$, but it also assumes that the evolution of the magnetic-field 
energy density (as well as the acceleration efficiency for the electrons) is proportional to the 
specific kinetic energy of the shock. A change in any of these (or other) assumptions of the model 
may affect our conclusions about the radial density profile of the CSM. 

We note however that the conclusions 
extracted in the following subsections are independent of the real value of $s$.

\subsubsection{Enhanced radial decline in the CSM density at late epochs}
\label{NOCSM}

Weiler et al. (\cite{Weiler2007}) found an enhancement of the flux-density 
decay rate after day $\sim$3100. These authors fitted the enhanced decay rate using 
an {\em ad hoc} exponential factor with an e-folding time of 
$\sim$1100 days, but noted that the same data can also be fitted 
(although worse) using a power-law decay with $\beta =$ $-$2.7. 

This decay takes place at all frequencies at the same time, leaving the 
spectral index of the supernova unaltered. These authors interpreted the 
exponential-like decay as being produced by a steeper, and fine-tuned, radial density 
profile of the CSM, beginning on day $\sim$3100 after the explosion. In this 
section, we show that even an extreme drop in the density of the CSM cannot explain the 
observed fall-off in the flux density at all frequencies.

If all the electrons accelerated during the supernova 
expansion (and not only those that have just been accelerated after being affected by the shock, 
as it can be derived from Sect. 5.3 of Weiler et al. \cite{Weiler2007}) contribute to the 
radio emission, a change in the CSM density profile is not enough to explain such a rapid 
decay of the flux density. Cooling effects, and eventually the escape of electrons from the 
emitting region, can have important effects on the evolution of the supernova flux density at 
these late epochs, which we now consider in more detail.

\begin{figure*}[!t]
\centering
\includegraphics[width=18cm]{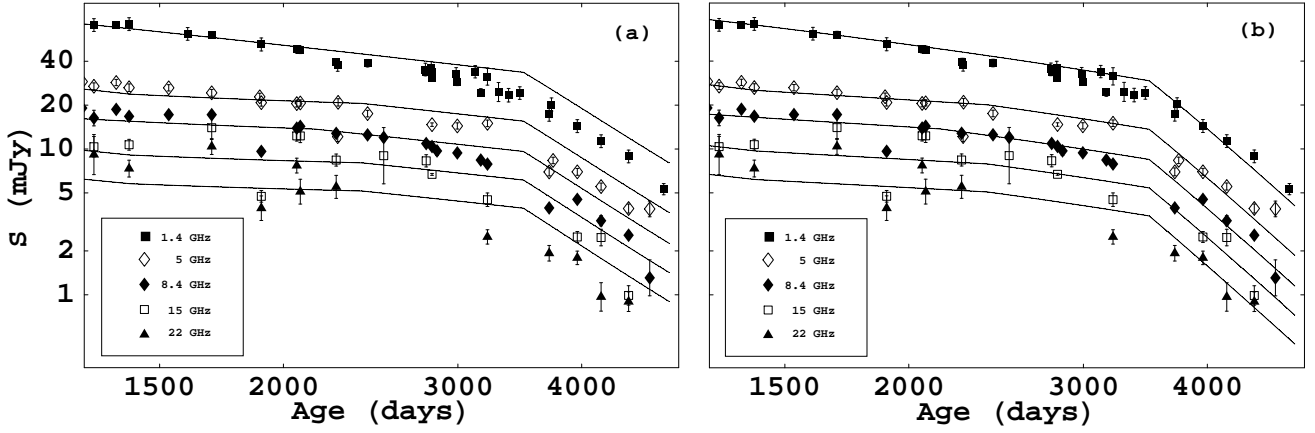}
\caption{Late radio light curves at several frequencies. Superimposed, RAMSES 
model obtained with a negligible CSM density after day 3100. (a) model with no escaping 
of the electrons from the emitting region. (b) model with a mean lifetime of the 
electrons inside the shell of 2500 days.}
\label{OnlyCooling}
\end{figure*}

We can obtain an upper bound to the flux-density decay rate after day 3100 by assuming 
that the CSM density profile approaches zero (or negligible values) from 
day 3100 onwards. In that case, the evolution of the expanding
structure will no longer be self-similar. However, we can still derive the
evolution of the expanding structure from the velocity fields given in Chevalier 
(\cite{Chevalier1982a}): the radial velocity 
of the shocked gas between the contact discontinuity and the forward shock is 
approximately equal to the expansion velocity of the contact discontinuity. If 
the supernova further expands into a negligibly dense CSM, the absolute 
(i.e., not fractional!) width of the shocked CSM region (i.e., from the contact 
discontinuity to the forward shock) will remain constant, and the expansion 
velocity of the forward shock will be roughly equal to the expansion velocity 
of the contact discontinuity. Using these assumptions, we can compute an upper 
bound to the flux-density decay corresponding only to electron cooling (i.e., turning 
off the electron-escape term). Thus, from day 3100 onwards, we 
ensure that the total number of relativistic electrons inside the shell remains constant. 
We also assume the ratio of the particle energy density to the magnetic-field 
energy density to be constant (as in the Chevalier model). 
In Fig. \ref{OnlyCooling}(a), we show the resulting simulated 
flux densities at late epochs superimposed on the observations reported in 
Weiler et al. (\cite{Weiler2007}). We note that the flux-density decay rate
predicted by RAMSES describes remarkably well the enhanced flux-density 
decay rate reported in Weiler et al. (\cite{Weiler2007}), although with systematic
slightly higher (around 10\%) flux densities. These systematics would be lower if we 
were to shift the initial epoch of negligible CSM to earlier dates by 50$-$100 days. In other 
words, {\em radiative-cooling effects, alone, are able to model the late flux-density decay 
rate of the radio light curves}, provided the CSM density becomes negligible from day 
3000$-$3100 onwards. A steep boundary in the CSM may be unrealistic, but it is conceivable 
that a rapid drop in the density is produced by the peculiarities in the onset of the stellar 
wind. 

For a smoother density drop, 
cooling effects cannot describe, alone, the observed flux-density decay rate. In that case, 
other contributions to the flux-density decay rate must be invoked. The escape of the electrons 
from the radiating region is a natural way to enhance the flux-density 
decay rate. For instance, if the CSM structure index, $s$, were changed from 2 to, say, 4 on day 
3100 (even disregarding the self-similar, standard interaction model scenario for radio 
supernovae), a mean lifetime of the electrons of about 1500 days (together with electron-cooling
effects) could also explain the observations. We note that the mean lifetime of the electrons 
affects the electron evolution during the entire expansion, not only after day 3100. 
In the case of a negligible CSM after day 3100, adding electron escape with a relatively 
long mean lifetime (around 2500 days) decreases the flux densities predicted by the 
model without increasing so much the enhanced decay rate at late epochs, resulting in a 
better fit to the data (see Fig. \ref{OnlyCooling}(b)). This actually corresponds to our final 
choice for 
the simultaneous modelling of the radio light curves and the expansion curve of SN\,1993J 
(see Sect. \ref{III} and Fig. \ref{SIMULFIT}).

It is difficult to identify a model with the right combination of both factors (electron 
escape and enhanced radial density profile of the CSM), since both quantities are completely
coupled. Therefore, the only clear conclusions we 
can reach at this point are: 1) an enhanced drop in the CSM density profile is needed to 
model the radio light curves after day 3100 and, depending on the amount of enhancement, 2) a 
finite mean lifetime of the electrons inside the radiating region may also be needed to explain 
the observations.

\subsubsection{Flux-density flare at $t \sim 100$\,days}

\begin{figure}[!t]
\centering
\includegraphics[width=9.5cm]{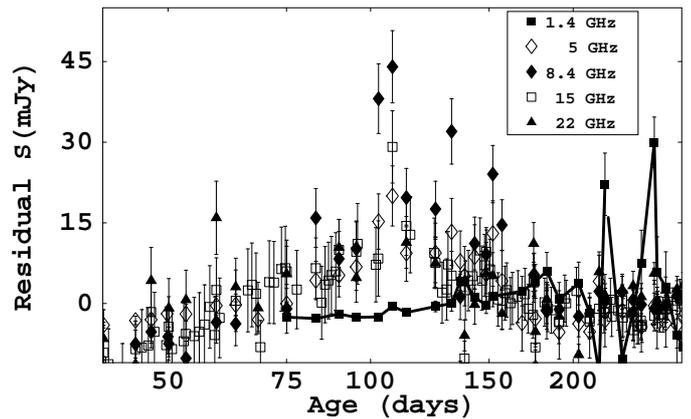}
\caption{Residual radio light curves from 40 to 250 days after explosion. The flux
densities at 1.4\,GHz are joined with lines for clarity.}
\label{FLARE}
\end{figure}

A zoom into Fig. \ref{SIMULFIT} around $t = 100$\,days uncovers a hint of excess emission from 
the supernova relative to the model predictions. This extra emission can also 
be seen in the residuals of the model reported in Weiler et al. (\cite{Weiler2007}). In Fig. 
\ref{FLARE}, we show the residuals around day 100 the after explosion. We note that the extra 
emission is detected at 5, 8.4, 15, and 22\,GHz, but not at 1.4\,GHz.
This extra emission of about $30 - 40$\,mJy (i.e., increase in total flux density of the supernova 
by $\sim 40$\%) lasted about 50 days. It was, therefore, a relatively large flux-density flare
of the supernova. It is remarkable that the flare was not detected at 1.4\,GHz. The source was 
optically thin at the other frequencies, so the intensity was roughly proportional to the electron 
number density. In contrast, according to our model, the source was optically thick to the 
synchrotron radiation at 1.4\,GHz during the flare. Thus, the intensity at that frequency was roughly equal to the 
source function ($j_{\nu} / \kappa_{\nu}$), which, avoiding the small effects of radiative cooling on the 
electron population, is independent of the electron number density (e.g. Pacholczyk \cite{Pacholczyk1970}). 
Therefore, a straightforward 
interpretation of Fig. \ref{FLARE} is that the number density of shocked CSM electrons suddenly 
increased by about 40\% around day 100 after the explosion (supernova radius of $\sim1150$\,AU). 
We note that an increase in 
the magnetic field when modelling the flare cannot explain
why it did not take place at 1.4\,GHz. Unfortunately, the supernova structure was not well 
resolved with VLBI at those early epochs and it is not possible to correlate the flare shown in Fig. 
\ref{FLARE} with any change in the emission structure of the supernova.

\begin{figure}[!t]
\centering
\includegraphics[width=9.5cm]{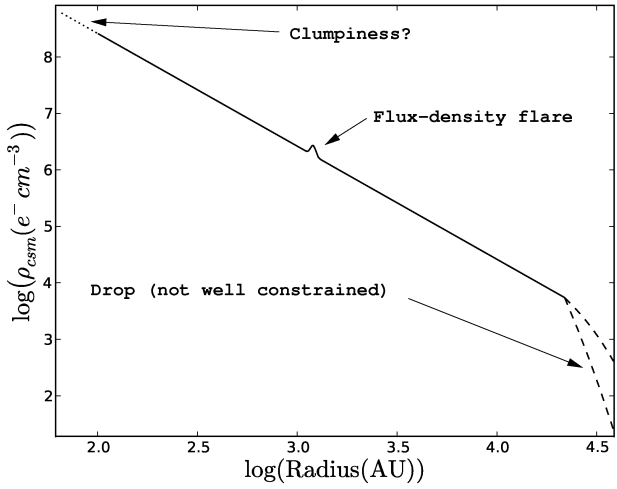}
\caption{Squematic representation of CSM density vs. distance to the explosion 
centre.}
\label{EsquemaCSM}
\end{figure}

In Fig. \ref{EsquemaCSM}, we show a schematic representation of the CSM radial
density profile inferred from the whole analysis reported here. On the one hand, for 
radial distances below $\sim$100\,AU there is evidence of clumpiness, based on 
the larger CSM opacities computed from the observations (see Fig. \ref{CSMTemp}). 
At a distance of $\sim$1150\,AU, a flare in the light-curve residuals provides evidence 
of an overdensity in the CSM. Finally, for radial distances larger than 21000\,AU, the 
CSM drops faster than $\propto r^{-2}$, the exact profile not being very well determined 
and dependent of the electron escaping used in the model.

\subsection{Evolution of the ejecta opacity}

\begin{figure*}[!t]
\centering
\includegraphics[width=18cm]{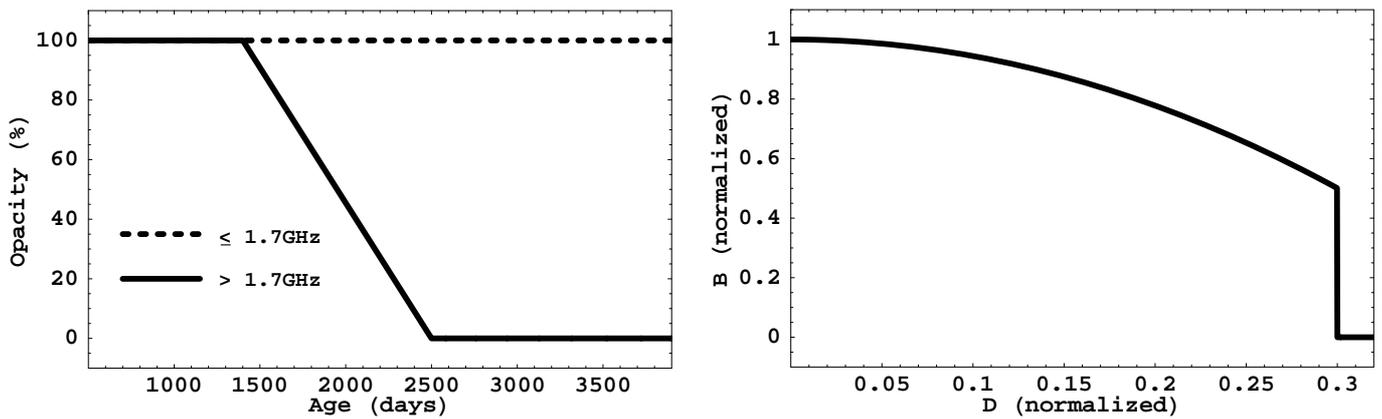}
\caption{{\em Ad hoc} assumptions of the RAMSES model, based in Marcaide et al. (\cite{Marcaide2009})
and Paper I. Left: evolution of the ejecta opacity 
used in the RAMSES simulations; 100\% means maximum opacity (total blockage of the radiation 
from behind the ejecta); 0\% means no ejecta opacity. Right: mean amplified magnetic
field (normalized to the magnetic field at the contact discontinuity) as a function of distance 
from the contact discontinuity (normalized to the shell radius).}
\label{Opac}
\end{figure*}

As proposed in Marcaide et al. (\cite{Marcaide2009}), a changing
ejecta opacity (which would be different for different frequencies) 
helps to explain the wavelength effects found in the expansion curve. In Paper I, we 
confirmed those wavelength effects. 
We implemented the opacity effects suggested in Marcaide et al. 
(\cite{Marcaide2009}) into the RAMSES model, although they are 
difficult to parametrize. Since the ejecta opacity can evolve in many 
different ways, several possibilities were tested. 
In Fig. \ref{Opac}, we show the final model selected for the ejecta 
opacity evolution. This model does 
not have any theoretical justification, but the true opacity should 
not be very different from the model 
proposed here. This model optimally fits all the radio data. In principle,
one would expect that the evolution of the ejecta opacity must not be the same 
at all frequencies higher than 1.7\,GHz. It seems more plausible that the 
opacity at higher frequencies should begin to decrease before those at lower 
frequencies. 
However, the data is not good enough to allow for such a careful modelling of 
the ejecta opacity.

\begin{figure}[!t]
\centering
\includegraphics[width=9.5cm]{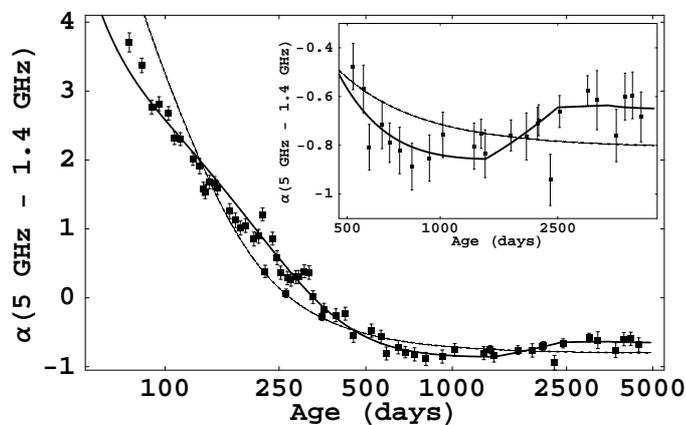}
\caption{Observed spectral indices between 1.4 and 5\,GHz reported by Weiler et al. 
(\cite{Weiler2007}). Dashed line, best-fit model reported by Weiler et al. (\cite{Weiler2007}). 
Continuous line, best-fit obtained with the RAMSES model. Inset is a blow up from day 
500 to 5000.}
\label{SPECIND}
\end{figure}

\begin{figure}[!t]
\centering
\includegraphics[width=9.5cm]{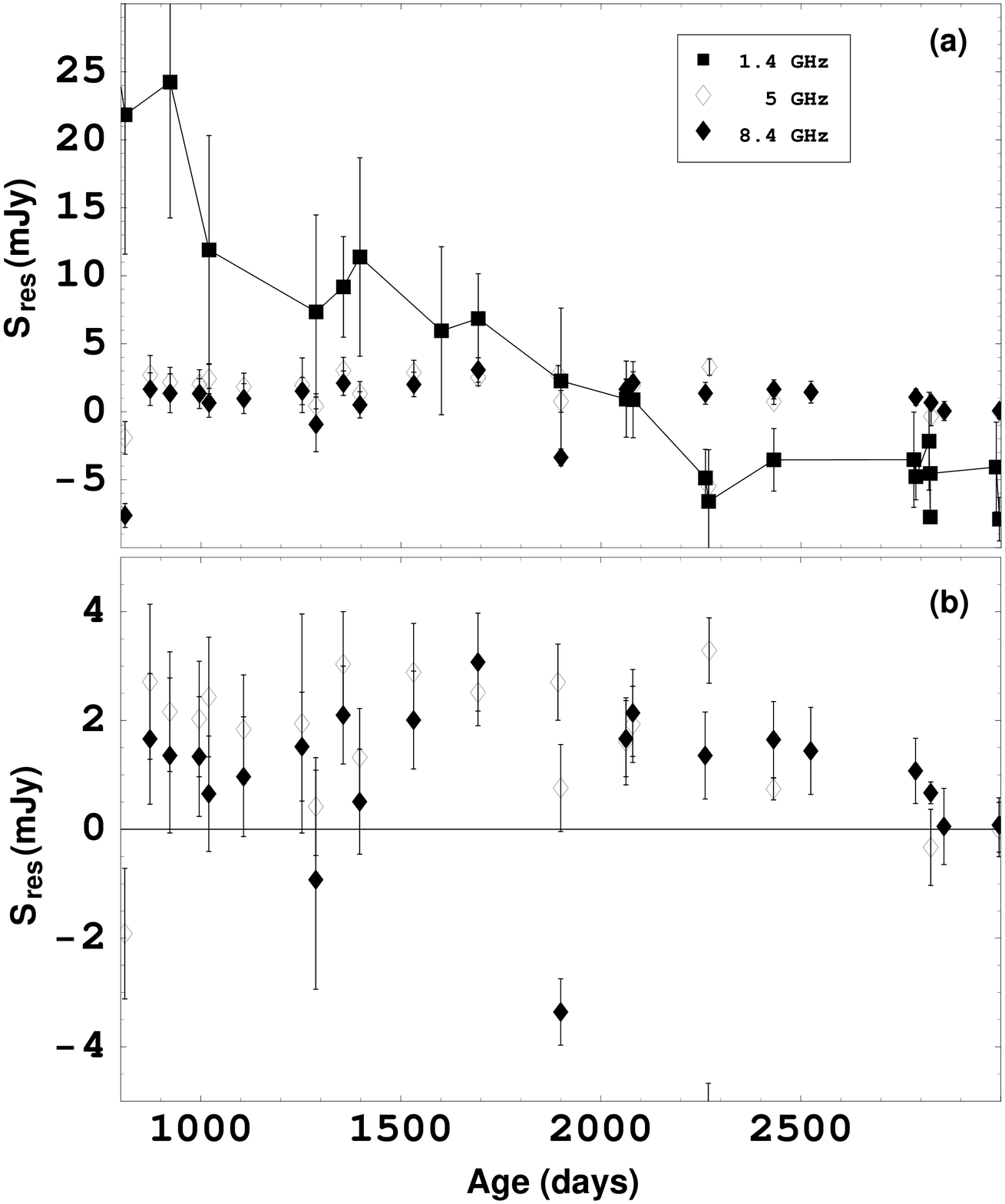}
\caption{Residuals of the SN\,1993J flux densities between days 1000 and 3000
after explosion, using the model of Weiler et al. (\cite{Weiler2007}): 
(a) all the residuals (1.4\,GHz 
residuals have been joined with a solid line for clarity); (b) zoomed, 
the residuals at 5 and 8.4\,GHz (the solid line now marks 
the zero mJy level).}
\label{WeilerRes}
\end{figure}

\begin{figure}[!t]
\centering
\includegraphics[width=8.5cm]{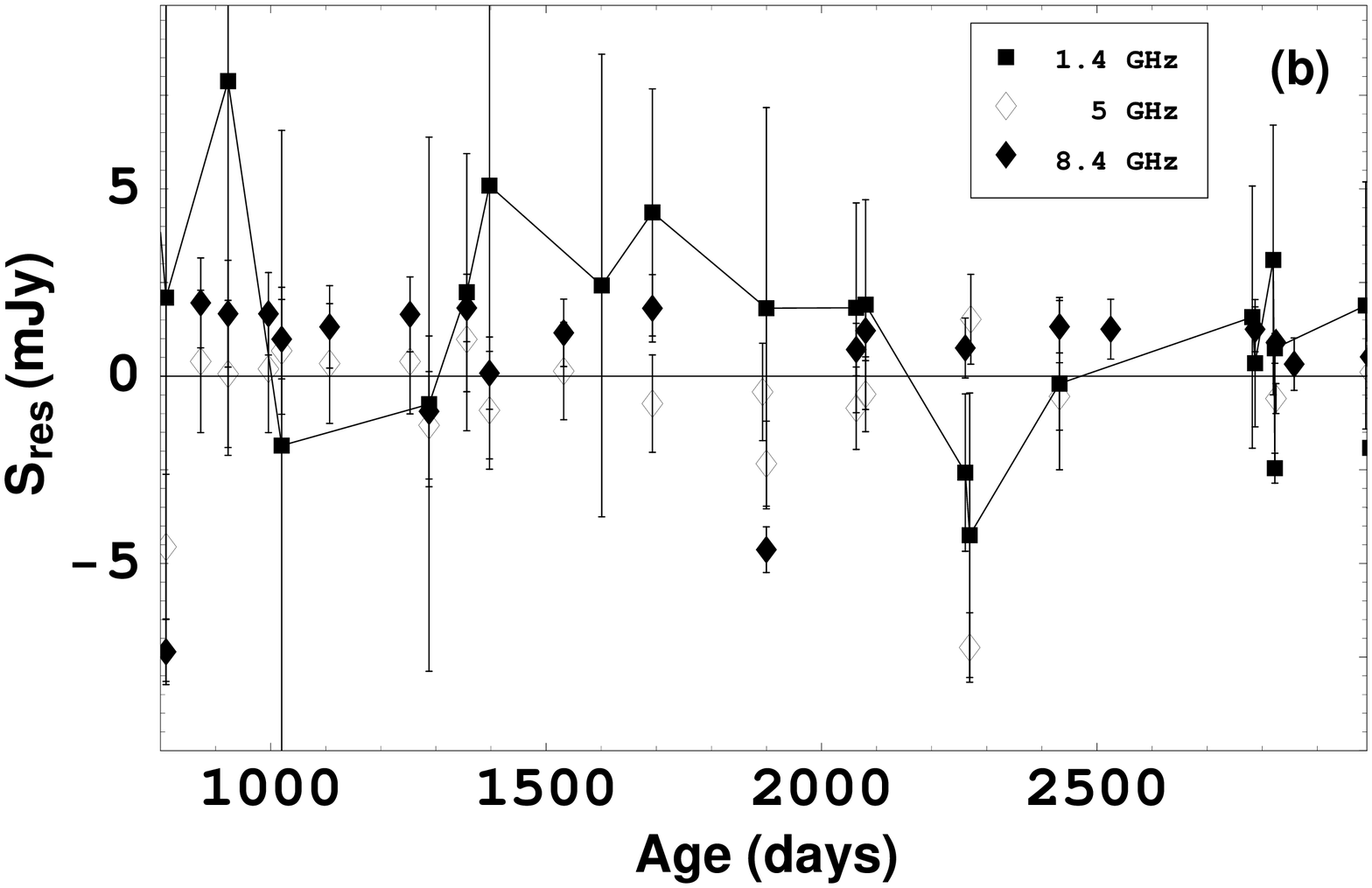}
\caption{Same as in Fig. \ref{WeilerRes} (a), but with the residuals
obtained with the RAMSES model.}
\label{RAMSESRes}
\end{figure}

The opacity at frequencies higher than 1.7\,GHz linearly decreases from
100\% (on day 1500) to 0\% (on day 2500), but remains constant at this frequency. 
This decrease in the ejecta opacity can explain the wavelength effects reported in the 
expansion curve of the supernova, but can also explain a slight 
increase in the SN\,1993J flux densities observed after day $\sim$1500 in 
the 8.4\,GHz data (and, to a lower degree, also in the 5\,GHz data) reported 
by Weiler et al. (\cite{Weiler2007}). 
In Fig. \ref{WeilerRes}, we show the 
flux-density residuals corresponding to the model used by Weiler et al. 
(\cite{Weiler2007}) around day 1500 after explosion. In Fig. \ref{RAMSESRes}, 
we show the residuals of RAMSES for the same time range. We note that the 
RAMSES residuals are typically half of those of 
Weiler et al. (\cite{Weiler2007}), and even $\sim$5 times smaller 
for some data points. The effect of opacity evolution can also be seen in the measured 
spectral indices. In Fig. \ref{SPECIND}, we show the spectral indices reported
by Weiler et al. (\cite{Weiler2007}) between 1.4 and 5\,GHz, comparing them with 
the model proposed by these authors and our model obtained with RAMSES. 
In the time window between 500 and 5000 days after the explosion, the systematic offsets 
between data and the model proposed by Weiler et al. can clearly be seen. Instead, the RAMSES 
model predicts remarkably well the evolution of the spectral index for all epochs, 
including the flattening beginning at an age $\sim$1000 days (P\'erez-Torres et al.
\cite{PerezTorres2002}; Bartel et al. \cite{Bartel2002}).

In Marcaide et al. (\cite{Marcaide2009}) and Paper I, we reported a fitted ejecta 
opacity of 80\% for all epochs and frequencies. This result might seem to 
be in conflict with the model proposed here for the evolution of the ejecta opacity. 
However, we also noted then that the ejecta opacities reported were too noisy for 
extracting any robust information about the possible evolution and/or spectral dependence 
of the ejecta opacity. Therefore, the value of 80\% reported there for the 
opacity should be taken as an approximate, average, value.

\subsection{Radial drop of the magnetic field}

Inside the shell it is also difficult to parametrize a radial drop in 
the magnetic field. After extensive testing,
we chose the following model to characterize the drop of $B$, 
as a function of distance, $D$, from the contact discontinuity 

\begin{equation}
B(D) = B_0(a\,D^2 + b\,D+c), 
\label{EqB}
\end{equation}

\noindent where the parameters $a$, 
$b$, and $c$ are chosen so that, for a fractional shell width of 0.3, 
$B_0$ is the mean magnetic field intensity of the shell and the
intensity at the forward shock is half the intensity at
the contact discontinuity (see Fig. \ref{Opac}). 

\begin{figure*}[!t]
\centering
\includegraphics[width=18cm]{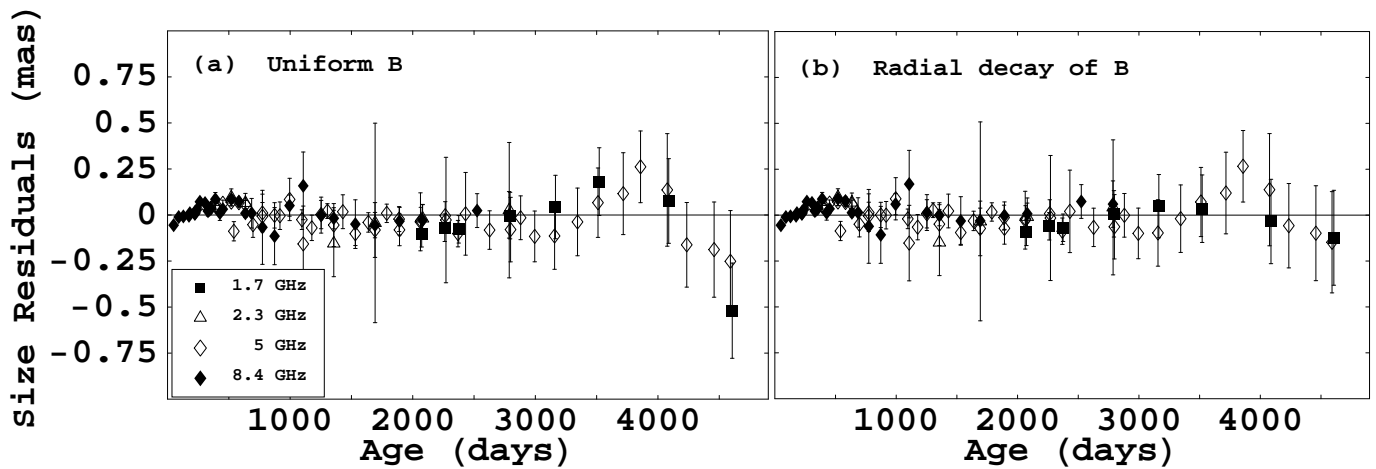}
\caption{Residuals of the expansion curve modelled with RAMSES using a uniform 
(a) and a radially-decaying (b) magnetic field inside the shell. The shape 
of the radial decay of the magnetic field is given by Eq. \ref{EqB} and shown 
in Fig. \ref{Opac}.}
\label{ResExpan}
\end{figure*}

Marcaide et al. (\cite{Marcaide2009}) pointed out that if the flux density
per unit beam decreases, the shell size estimate will be biased towards a smaller
value, provided the magnetic field drops radially in the shell (see their Sect. 
7.1.2). Thus, the exponential-like decrease in flux densities after day 3100, combined
with a magnetic field structure similar to that given in Eq. \ref{EqB}, should translate 
into progressively biased estimates of the shell size, and therefore an increase 
in the (observed) deceleration of the expansion curve. If we include 
radial drops in the magnetic field steeper than that corresponding to Eq. 
\ref{EqB} (such as a linear or a concave-like decay), we obtain poorer 
fits to the expansion curve. Therefore, we conclude that the radial drop 
in the magnetic field inside the shell must be smooth. In Fig. \ref{ResExpan}, we compare
the expansion-curve residuals obtained with a uniform magnetic
field inside the shell (a) with those for the magnetic-field structure given by Eq. 
\ref{EqB} and shown in Fig. \ref{Opac}. It can be seen that the residuals for the 
latest VLBI epochs are more accurately fitted using the radially-decaying magnetic field. 

We note however, that the conclusions extracted from this section are based on noisy
images (from the latest epochs at which the supernova could be barely imaged).
Therefore, these conclusions should be considered with caution.

\section{Conclusions}
\label{V}

We have developed software to simultaneously model the VLBI 
expansion curve and the radio light curves of supernova SN\,1993J.
This software takes into consideration the 
evolution of the magnetic field energy density and the hydrodynamic 
evolution of the expanding shock, as well as the relativistic 
acceleration of CSM electrons as they interact with the 
supernova shock. All these processes have been implemented following 
Chevalier (\cite{Chevalier1982a},\cite{Chevalier1982b}). 
Our software also accounts for the radiative cooling of the electrons, as well as 
SSA and inverse Compton scattering. The escape of electrons from the 
radiating region is also considered.

We have modelled the whole radio-data set of 
SN\,1993J (radio light curves and expansion curve) using one 
single model. We have 
considered a changing opacity of the supernova ejecta to the 
radio emission and a radial decay in the magnetic field
within the radiating region. The structure index of the CSM is set to 
$s= 2$ up until day 3100 after the explosion. From that day onwards, a higher 
value of $s$, or even a negligibly dense CSM, is required to model the radio 
light curves. In the case of a negligibly dense CSM, cooling
effects, alone, are able to predict the exponential-like 
flux-density drop reported by Weiler et al. (\cite{Weiler2007}).
When the CSM is not negligible after day 3100, 
a finite mean lifetime of the electrons inside the emitting 
region is also needed to model the late radio light curves.
We also find, at all frequencies apart from 1.4\,GHz, an unmodelled increase in 
the flux density of the supernova around day 100 after the explosion. This 
increase represents $\sim 40$\% of the total flux 
density of the supernova at these epochs. We suggest that this sudden 
increase in the flux density may be caused by an increase in the density of 
the shocked CSM electrons around day 100.

In our model, the ejecta opacity remains constant at 1.7\,GHz, but changes 
at the other frequencies from 100\% (at day 
1500 after explosion) to 0\% (at day 2500). This evolution of the 
opacity explains the effects found in the expansion curve and is also able 
to explain some effects found in the radio light 
curves (hence, in the evolution of the spectral indices).

We also found evidence of a radial drop in the magnetic field inside the 
radio shell. When combined with the enhanced flux-density decay rate of the radio 
light curves at late epochs, this drop explains the enhanced deceleration found in 
the expansion curve of the latest VLBI observations at all frequencies.

\acknowledgements{ IMV is a fellow of the Alexander von 
Humboldt Foundation in Germany.
The National Radio Astronomy Observatory is a facility of the
National Science Foundation operated under cooperative agreement
by Associated Universities, Inc. The European VLBI Network is a
joint facility for European,Chinese, South African and other radio
astronomy institutes funded by their national research councils.
Partial support from Spanish grant AYA~2005-08561-C03-02 is
ackwnowledged. AA and MAPT acknowledge support by the Spanish 
Ministry of Education and Science (MEC) through grant AYA 
2006-14986-C02-01, and by the Consejer\'{\i}a de Innovaci\'on, 
Ciencia y Empresa of Junta de Andaluc\'{\i}a through grants 
FQM-1747 and TIC-126.}

\appendix

\begin{onecolumn}

\section{The RAMSES model}
\label{RAMSESApp}

\subsection{Equation of the electron evolution}

As implemented by Fransson \& Bj\"ornsson (\cite{Fransson1998}), we 
simulated the relativistic electron population using the equation of 
continuity in energy space

\begin{equation}
{\dot N (E,t)} = -{\nabla_E}(\dot E N(E,t)) + S(E,t) - L(E,t)~,
\label{EqEvol}
\end{equation}

\noindent where $N(E,t)$ is the number of electrons that fill the radiating region 
with energies between $E$ and $E + dE$ at time $t$, $S(E,t)$ is the source function 
accounting for the continuum acceleration of CSM electrons by the shock, $L(E,t)$
takes into account the possible escaping of electrons from the emitting region, and 
$\dot E$ represents the energy loss (or gain) of an electron with energy $E$ at 
time $t$. In the following subsections, we explain each term of Equation \ref{EqEvol}
in more detail.

\subsection{Energy losses of the electrons}

Our software considers several terms for the computation of energy losses of the
relativistic electrons (negative terms for $\dot E$). These terms are the expansion 
of the relativistic gas (which results in $\dot E \propto E/t$) and radiative 
processes (synchrotron, Coulomb, and inverse Compton; see 
Pacholczyk \cite{Pacholczyk1970}). 

In the case of 
synchrotron losses, we have $\dot E \propto \bar{B_{\perp}^2} E^2$, where 
$\bar{B_{\perp}^2}$ is the mean of the square of the perpendicular component of 
the magnetic field to the (random) trajectories of the electrons. 

Coulomb losses are modelled with 
$\dot E \propto (f_a + \log{E}) E$, for free-free processes, and with 
$\dot E \propto (f_b + \log{E})$, for processes related to the ionization of the 
medium ($f_a$ and $f_b$ are constants that can be computed from the density of 
hydrogen in the medium; see Pacholczyk \cite{Pacholczyk1970}).

We also included losses due to the inverse Compton effect, which only affects the
earliest radio light curves at the highest frequencies. In this case, 
$\dot E \propto J~ E^2$, where $J$ is the photon density inside the radiating region.
The photon-density estimates used in our modelling were taken from Fransson \&
Bj\"ornsson (\cite{Fransson1998}) (see their Fig. 6).

\subsection{Energy gain of the electrons}

Once the electrons have been accelerated by the shock, their main process of energy
gain is SSA, since the probability of being re-accelerated is very low (the 
acceleration efficiency of the shock $\sim10^{-5}$, see fitted value of $F_{rel}$
in Table \ref{RAMSESTable}). To compute the effect of self-absorption in the electron 
energy distribution, we must take into account possible radial 
variations in the amplitude of the magnetic field in the shell, especially for the 
emission frequencies for which the optical depth is close to 1. It can be shown
(see Mart\'i-Vidal \cite{MartiVidal2008}) that the expression to use in this case is

\begin{equation}
\left(\dot E N \right)_{ssa} = E^2 \nabla_E \left(\frac{n(E)}{E^2}\right) 
\int{\int{\frac{K_s}{\nu^2} I_{\nu}(S) F(x)\,dS}\,d\nu}~~,
\label{SelfAbs}
\end{equation} 

\noindent where the integral over $S$ is the integral over the radiating region, and
the integral over $\nu$ is that over all the synchrotron emission frequencies. 
$I_{\nu}(S)$ is the synchrotron emission intensity at the point $S$ of the source and
at frequency $\nu$. The factor $n(E)$ is the density of electrons with energies between 
$E$ and $E + dE$ in the point $S$ of the source. This equation is a generalization of 
that used by Fransson \& Bj\"ornsson (\cite{Fransson1998}), but 
can be applied to magnetic fields with generic amplitude distributions inside 
the source. The amplitude of the magnetic field, as a function of its position $S$, 
appears implicitly in $x$ and $K_s$ (the definitions of $x$, $K_s$, and $F(x)$ can be 
seen in Pacholczyk \cite{Pacholczyk1970}).

\subsection{Source of electrons}
\label{SOURCE}

The source function of electrons depends on energy and time. We assumed 
an expression similar to that one used by Fransson \& Bj\"ornsson 
(\cite{Fransson1998})

$$S(E,t) = F_{rel} F_{nor} N_0 (R/R_0)^{-s} (V/V_0)^2 E^{-p}~,$$

\noindent where $F_{rel}$ is the fraction of electrons of the recently-shocked CSM that
have been accelerated, $N_0$ is the electron density at a reference epoch (when the 
radius of the supernova shell was $R_0$ and the expansion velocity was $V_0$), $R$ is 
the radius of the shell, $V$ its expansion velocity, and $F_{nor}$ is a normalization 
factor. This normalization factor scales the amplitude of an electron population 
distribution given by $N(E) \propto F_{rel} N_0$ into a population of relativistic 
electrons with an energy distribution given by $N(E) \propto E^{-p}$, in such a way 
that the number of electrons is conserved.

The factor $(V/V_0)^2$ is included to make the acceleration efficiency of the shock
proportional to its specific kinetic energy. This proportionality has been found to 
more accurately describe the electron acceleration, according to Fransson \& Bj\"ornsson 
(\cite{Fransson1998}).

\subsection{Escaping of electrons from the emitting region}

Electrons could escape from the emitting shell, either towards the unshocked CSM or 
into the region of expanding ejecta.
The escaping of electrons from the emitting region was assumed to be independent
of the electron energies. All the electrons have velocities close to the speed of light, 
and the effect of the magnetic-field pressure inside the emitting region should be 
similar for all the electrons. These two reasons make the probability of escaping from 
the shell roughly equal for all electrons. Therefore, the expression used to model 
the escaping of electrons is 

\begin{equation}
L(E,t) = N(E,t)/t_m~~, 
\label{EqEsc}
\end{equation}

\noindent where $t_m$ is the mean lifetime of an electron inside the radiating shell. 

\subsection{Electron evolution and radio emission}

Numerical integration of Eq. \ref{EqEvol} is performed semi-implicitly 
(see Mart\'i-Vidal \cite{MartiVidal2008}). This equation can be written as 
$ \dot N = f(N,\nabla_E N, E, t)$, where $f$ is a functional that depends on $N(E)$, $E$, 
and $t$. If we apply a binning of $N(E,t)$ in energy and time\footnote{We apply 
a logarithmic binning in energy and time to optimize the accuracy of the 
numerical integration}, such that $N_{k,i}$ is the electron number at energy 
$E_k$ and time $t_i$, we can approximate Eq. \ref{EqEvol} by

\begin{equation}
\frac{N_{k,i+1} - N_{k,i}}{t_{i+1} - t_i} = 
\frac{1}{2}\left(f + (N_{k,i+1} - N_{k,i}) \nabla_N f \right).
\label{TimeEvol}
\end{equation}  

Knowing the electron population at time $t_i$ (i.e., knowing $N_{k,i}$ for all $k$), 
it is possible to compute the new population at time $t_{i+1}$ (i.e., $N_{k,i+1}$ 
for all $k$). However, care must be taken when computing $\dot E N(E,t)$ in Eq. 
\ref{EqEvol}, especially the term related to SSA, for which 
an accurate estimate of the intensity distribution of synchrotron radiation inside 
the shell is needed.

Once $N_{k,i+1}$ was obtained, we computed the emission intensity at all the 
desired frequencies, and transmitted the result through a filter that takes into account
the opacity due to the unshocked CSM electrons. This opacity can be estimated 
using the equation (e.g., P\'erez-Torres \cite{MATesis})

\begin{equation}
\tau_{\nu} = C_f^{-2}\,\frac{0.17}{\nu^2}\,
\int_{R}^{\infty}{n_{cs}(r)^{2}\,T(r)^{-3/2}\,
\left(1 + 0.13\,\log\left(\frac{T(r)^{3/2}}{\nu}\right)\right)\,dr} 
\label{ThermElec}
\end{equation}

\noindent given in the cgs system, where $C_f$ is the compressing factor of the shocked
CSM density relative to the unshocked CSM density. This factor is $\sim$4 (e.g., 
Dyson \& Williams \cite{Dyson1980}),
$T(r)$ is the temperature of the unshocked CSM medium as a function of distance to 
the supernova explosion centre. RAMSES uses the same temperature profile used by 
Fransson \& Bj\"ornsson (\cite{Fransson1998})

\begin{equation}
T(r) = \textrm{Max} \left(T_l\times\left(\frac{R_0}{r}\right)^{\delta}, T_c\right)~~,
\label{TempMod}
\end{equation}

\noindent where $T_c = 2\times10^5\,K$, $\delta = 1$, and $T_l$ is a fitting 
parameter.

Afterwards, the opacity-corrected flux densities were used to generate the 
synthetic radio light curves and the VLBI images. A cut-off was then applied to the
VLBI images according to the sensitivity of the interferometric arrays. Finally, the
Common-Point method (see Marcaide et al. \cite{Marcaide2009} and Mart\'i-Vidal
\cite{MartiVidal2008}) was applied to generate the synthetic expansion
curve at all frequencies.

\subsection{Fitting procedure}

When the synthetic radio light curves and expansion curve have been generated,
the software compares them with the observations and computes the corresponding 
$\chi^2$. Then, new values of the simulation parameters are computed to minimize
the $\chi^2$. This step is performed using the SIMPLEX algorithm (e.g., Nesa 
\& Coppins \cite{SIMPLEX}). The whole process is iterated until convergence of 
SIMPLEX.

\subsection{Test of RAMSES}

The integration code of RAMSES was tested in several ways. The conservation
of the number of electrons after applying Eq. \ref{TimeEvol} was 
tested. On the one hand, at low energies ($E \rightarrow m c^2$) the particle 
number is not well conserved, since $\dot E N(E,t)$ was computed using expressions 
that are only valid in the relativistic regime. However, the low-energy region 
of the electron population is full of thermal (i.e. non-accelerated) electrons, 
which dominate the distribution and, moreover, those electrons do not 
contribute to the synchrotron emission. Therefore, this lack of conservation of 
the non-relativistic electron population is not crucial in our simulations.
On the other hand, the upper limit to the energy 
distribution used in the simulations ($10^4\,m c^2$) produces a small 
generation of electrons at the highest-energy boundary of our integration 
window. Nevertheless, this effect is very small (of the order of $10^{-6}$ 
times the source function $S(E,t)$ at those energies).
We also tested the degree of convergence of the solutions as a function
of the number of bins in energy and time spaces. Setting 3000 time steps (between
days 2 and 4900 after explosion) and 1000 bins in energy space (between 1 and 
$10^4$ Lorentz factors), we obtain results that differ only $\sim 0.1\%$ 
from the results obtained using twice the number of bins.

The electron population obtained also evolves as it is 
expected if we consider the energy gains and losses, as well as the 
injection and escaping of electrons. In Fig. \ref{ExampleRAMSES},
we show an example of the evolution of an electron population obtained after 
a RAMSES run. At 10 days after explosion, a small {\em bump} is seen at 
$E \sim 10\,m\,c^2$, due to the energy gain produced by SSA 
(i.e., electrons with this energy emit with a peak frequency 
for which the mean opacity is $\tau \sim 1$). Later, radiative losses dominate,
and the spectral index tends to increase 1 (in absolute value) at higher 
frequencies and decrease 1 (in absolute value) at lower frequencies. Both 
effects can be appreciated in Fig. \ref{ExampleRAMSES}, with the help of 
the straight line also shown in the figure. 

\begin{figure}[!h]
\centering
\includegraphics[width=9.5cm]{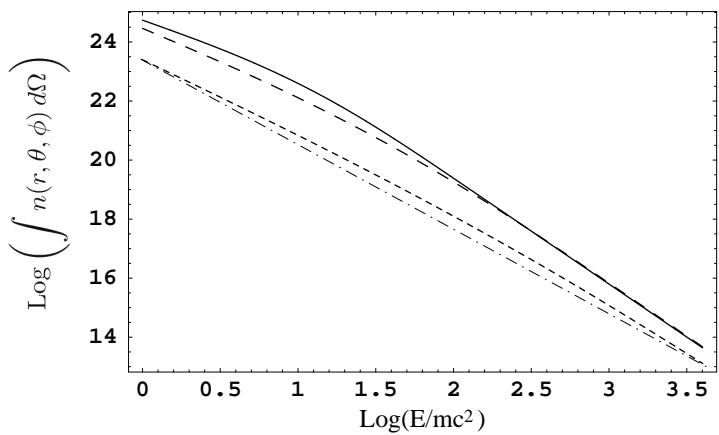}
\caption{Energy distribution of the solid-angle integral of an electron 
population simulated with RAMSES, computed in the forward shock. The continuum
line is the population at 10 days after explosion. The long-dashed line is the 
population at 100 days. The short-dashed line is the population at 1000 days.
The dot-dashed line is a straight line, plotted for visual aid.}
\label{ExampleRAMSES}
\end{figure}

\end{onecolumn}

\end{document}